\documentclass[
aps,
prapplied,
superscriptaddress,
twocolumn,
notitlepage
]{revtex4-2}
\usepackage{
graphicx,amsmath,mathtools,amsfonts,bm
}
\usepackage[normalem]{ulem}
\usepackage[usenames, dvipsnames]{xcolor}

\usepackage{hyperref}
\hypersetup{colorlinks=true, linkcolor=blue, citecolor=blue, urlcolor=blue}

\begin{document}
\title{Ferroelectric transmon}

\author{M. Donaire}
\affiliation{Departamento de F\'isica Te\'orica, At\'omica y \'Optica and IMUVA,  Universidad de Valladolid, Paseo Bel\'en 7, 47011 Valladolid, Spain}
\affiliation{
Univ. Grenoble Alpes, CNRS, Grenoble INP, Institut Néel, 25 Rue des Martyrs, 38042, Grenoble, France
}
\author{A. Cano}
\affiliation{
Univ. Grenoble Alpes, CNRS, Grenoble INP, Institut Néel, 25 Rue des Martyrs, 38042, Grenoble, France
}
\begin{abstract}
Superconducting qubits are a leading platform for quantum computing. However, simultaneously achieving low noise sensitivity to suppress decoherence and sufficient anharmonicity to enable fast gate operations remains a central challenge. Here, we introduce the concept of the ferroelectric transmon (FEmon), in which the Josephson junction is shunted by a ferroelectric, or incipient ferroelectric, capacitor. We show, in particular, that the nonlinear ferroelectric response of the capacitor provides an additional degree of freedom for optimizing qubit anharmonicity while preserving operation in the charge-noise-insensitive regime.
\end{abstract}

\date{\today}
\maketitle

\section{Introduction}

In the field of quantum computation, the transmon has emerged as the workhorse superconducting qubit. It serves not only as the fundamental building block of arguably the most advanced quantum computing technologies~\cite{arute2019quantum,vandamme2024}, but also as a versatile platform for exploring fundamental quantum phenomena~\cite{leib2010bose}.
The transmon consists of a Josephson junction shunted by a capacitor, forming a simple nonlinear LC circuit~\cite{schoelkopf07-pra,wendin17}. In the large-capacitance regime, the sensitivity of the qubit transition energy to charge noise is exponentially suppressed. However, this robustness comes at the cost of a substantial reduction in anharmonicity. The resulting decrease in level spacing increases the probability of leakage beyond the computational subspace, thereby limiting both gate fidelity and the ultimate speed of quantum operations. These constraints motivate the search for new superconducting qubit architectures that combine strong charge-noise insensitivity with enhanced anharmonicity (see, e.g., Refs.~\cite{wendin17,bao2021fluxonium,hyyppa2022unimon}).

In this context, we introduce the new concept of the ferroelectric transmon (FEmon). In this superconducting qubit architecture, the charging energy is governed by the response of a ferroelectric, or incipient ferroelectric, capacitor. As we show below, the nonlinear dielectric response of this element provides an additional degree of freedom for simultaneously engineering the qubit spectrum and controlling the transition frequencies. This added tunability enhances the selective addressability of individual transitions while preserving the exponential insensitivity of the qubit to charge noise.

\begin{figure}[b!]
\includegraphics[height=.225\textwidth]{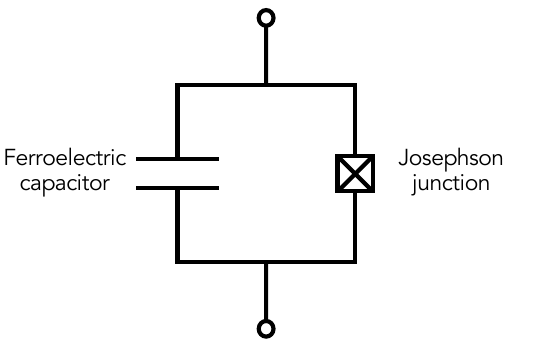}
\caption{\label{f:transmon}Effective circuit of the ferroelectric transmon (FEmon). The Josephson junction is shunted by a ferroelectric capacitor, in either its incipient ferroelectric (paraelectric) state or ferroelectric one. }
\end{figure}

To illustrate the FEmon concept, we consider the equivalent-circuit Hamiltonian,
\begin{align}
\hat H = \frac{a}{2}(\hat n - n_g)^2 + \frac{b}{4}(\hat n - n_g)^4 - E_J \cos \hat \phi .
\label{HH}
\end{align}
Here, $\hat n$ denotes the charge on the capacitor, expressed in units of $2e$ (one Cooper pair), while $\hat \phi$ is the superconducting phase difference across the Josephson junction (see Fig.~\ref{f:transmon}). These variables are canonically conjugate, and the corresponding operators satisfy the commutation relation $[\hat \phi,\hat n]= i$. In the phase representation, the charge operator is given by $\hat n = -i\partial_{\phi}$. The quantity $n_g$ represents the offset charge induced by the electromagnetic environment.

The first two terms of the FEmon Hamiltonian in Eq.~\eqref{HH} originate from the ferroelectric shunt capacitance and define the overall charging energy $E_C$, as discussed below. The cosine term, in turn, arises from the Josephson junction, with $E_J$ denoting the corresponding Josephson energy. The Hamiltonian in Eq.~\eqref{HH} differs from that of the standard transmon in that the large-capacitance regime is described more realistically through the inclusion of the $\mathcal{O}(\hat n^4)$ term. This formulation not only enables a consistent treatment of the entire transmon regime, including the $a \to 0^+$ limit, but also reveals a distinct operating regime in which the capacitor enters a ferroelectric state characterized by spontaneous polarization (see, e.g., Ref.~\cite{levanyuk24}). In the following, we refer to these two regimes as the incipient ferroelectric and ferroelectric regimes, respectively. Equation~\eqref{HH} provides a unified description of both by treating $a$ as a control parameter that approaches zero and eventually changes sign, from positive values in the incipient ferroelectric (paraelectric) regime to negative values in the ferroelectric regime.

We note that the FEmon concept implies that the nonlinear response of the capacitor arises primarily from the aforementioned $a \to 0$ behavior rather than from the magnitude of the higher-order charge term itself. In other words, the material-specific parameter $b$ in Eq.~\eqref{HH} need not be particularly large to produce a substantial nonlinear effect. Indeed, several ferroelectric materials lie close to the so-called tricritical point, at which the phase transition changes from first order to second order. At this point, $b=0$, and the inclusion of a higher-order $\mathcal{O}(\hat n^6)$ term becomes necessary. In practice, the material parameters of ferroelectric capacitors can be tuned through chemical composition, strain, or temperature, for example. More generally, the ferroelectric criticallity ($a \to 0$), together with relatively weak higher-order terms, further implies that the regime $E_J/E_C \gg 1$ is the natural operating regime of the FEmon qubit.

Further, from a methodological perspective, we note that the excited states of the FEmon Hamiltonian in Eq.~\eqref{HH} rapidly become dominated by the quartic term in the charge sector. This means that the highly excited eigenstates are well approximated by charge eigenstates and therefore correspond to plane waves in the phase representation,
$\psi_m(\phi) \equiv \frac{1}{\sqrt{2\pi}} e^{i m \phi}$.
Accordingly, the corresponding energy levels approach
$E_m \approx \frac{a}{2}m^2 + \frac{b}{4}m^4$,
whenever $E_m \gtrsim E_J$. These plane-wave states therefore provide a convenient basis for treating the FEmon Hamiltonian, whereas the use of Mathieu functions---i.e., the eigenstates of the standard transmon---introduces unnecessary complications in the general case. In particular, in the absence of offset charge, it is sufficient to retain a number of charge eigenstates of order $\max[(4E_J/b)^{1/4},(2E_J/a)^{1/2}]$ to accurately diagonalize the Hamiltonian in Eq.~\eqref{HH}.

\section{Incipient ferroelectric regime}

\begin{figure*}[t!]

\begin{tabular}{r c l}
\vspace{-43em}\\
\footnotesize $\displaystyle \frac{a}{b} $ & $=$ & $10$ \\
\vspace{7em}\\
 \footnotesize $\displaystyle \frac{a}{b} $    & $=$ & $1$\\ 
\vspace{7.em}\\
\footnotesize $\displaystyle \frac{a}{b} $ & $=$ & $0.1$\\
\end{tabular}
\includegraphics[width=.9\textwidth]{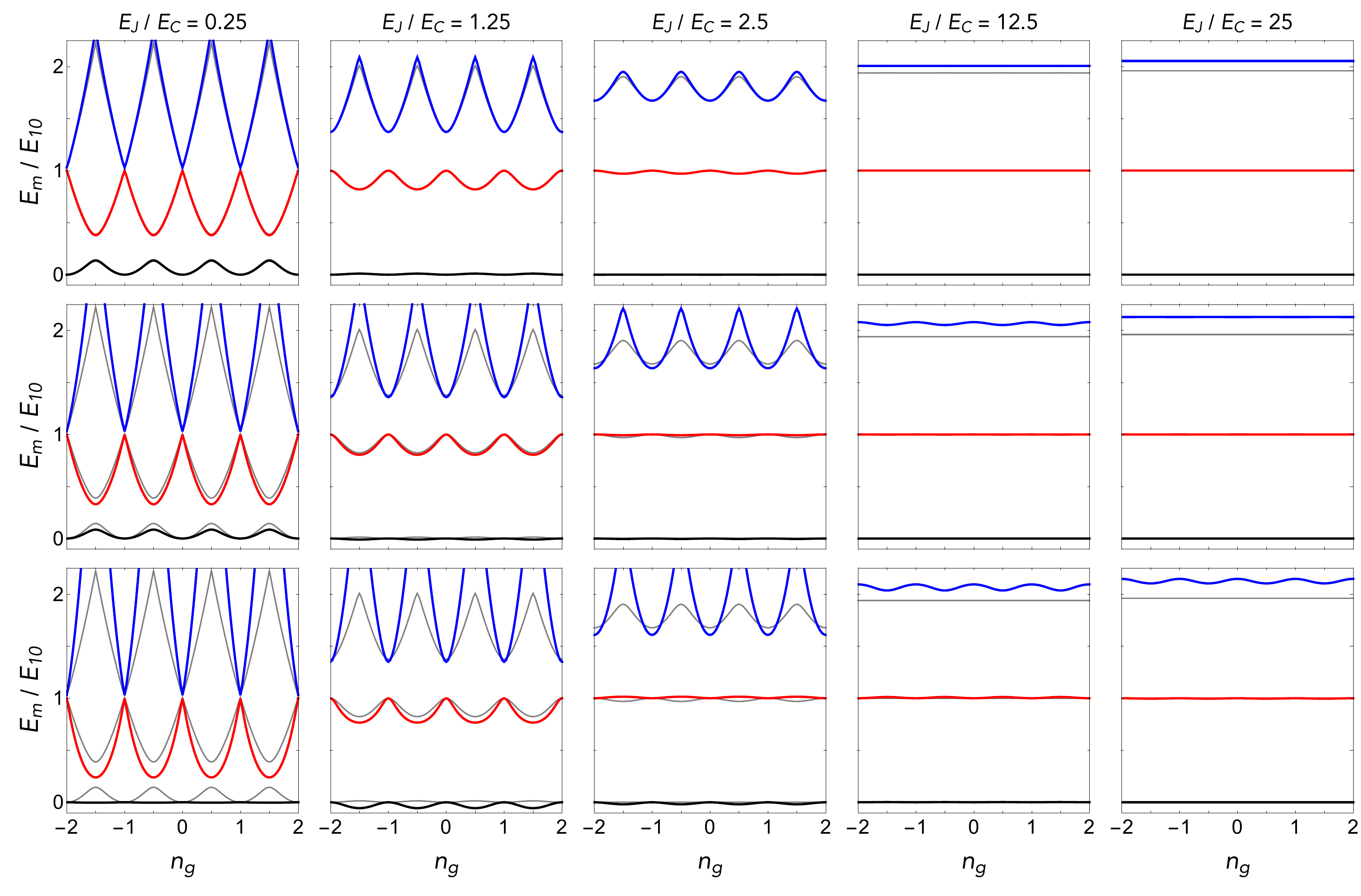} 
\vspace{-1em}
\caption{
Low-lying energy levels $E_m$ ($m=0,1,2$, shown in black, red, and blue, respectively) of the FEmon in the incipient ferroelectric regime, compared with the corresponding transmon levels (gray), as a function of the effective offset charge $n_g$ for different values of $E_J/E_C$ and $a/b$. 
The energies are expressed in units of the transition energy $E_{10}$ evaluated at $n_g=0$. Increasing $E_J/E_C$ (from left to right) progressively suppresses the charge dispersion, while decreasing $a/b$ (from top to bottom) enhances the anharmonicity.
}
\label{f:levels}
\end{figure*}

To illustrate the enhanced tunability of the FEmon relative to the standard transmon, we first analyze its operation in the incipient ferroelectric regime. That is, we first consider the large-capacitance limit $a \to 0^+$ in Eq.~\eqref{HH} describing the tendency toward the emergence of a spontaneous polarization in the capacitor, while this polarization itself remains zero. Figure~\ref{f:levels} illustrates the low-lying spectrum of the FEmon in this regime as a function of the offset charge $n_g$ for different values of the ratio $E_J/E_C$. Here, the charging energy is defined as
$E_C = \frac{a}{2} + \frac{b}{4}$.

The top panels, calculated for $a/b = 10$, show the behavior obtained when the shunt capacitor remains relatively far from the ferroelectric transition. As can be seen, the sensitivity to charge noise can be strongly suppressed by increasing $E_J/E_C$, as in the standard transmon. However, the anharmonicity evolves differently and is visibly enhanced at large values of $E_J/E_C$.

Furthermore, the overall anharmonicity of the FEmon undergoes a substantial enhancement as the ratio $a/b$ decreases, i.e., as the shunt capacitor is driven closer to the ferroelectric transition point. This effect is already apparent for $E_J/E_C = 1.25$, corresponding to the charge-qubit or quantronium regime rather than the transmon regime. We note that the enhancement is particularly pronounced at the charge-degeneracy point $n_g = 1/2$, where the charge dispersion of the first two levels is flat to leading order. Moreover, the enhanced anharmonicity of the FEmon relative to the standard transmon persists upon further increasing $E_J/E_C$. This demonstrates that the FEmon allows simultaneous optimization of two key qubit characteristics already within its incipient ferroelectric regime.

\begin{figure*}[t!]
\flushleft \hspace{9em} (a) \hspace{17em} (b)
\\[-1.5em]
\centering
\includegraphics[width=.3\textwidth]
{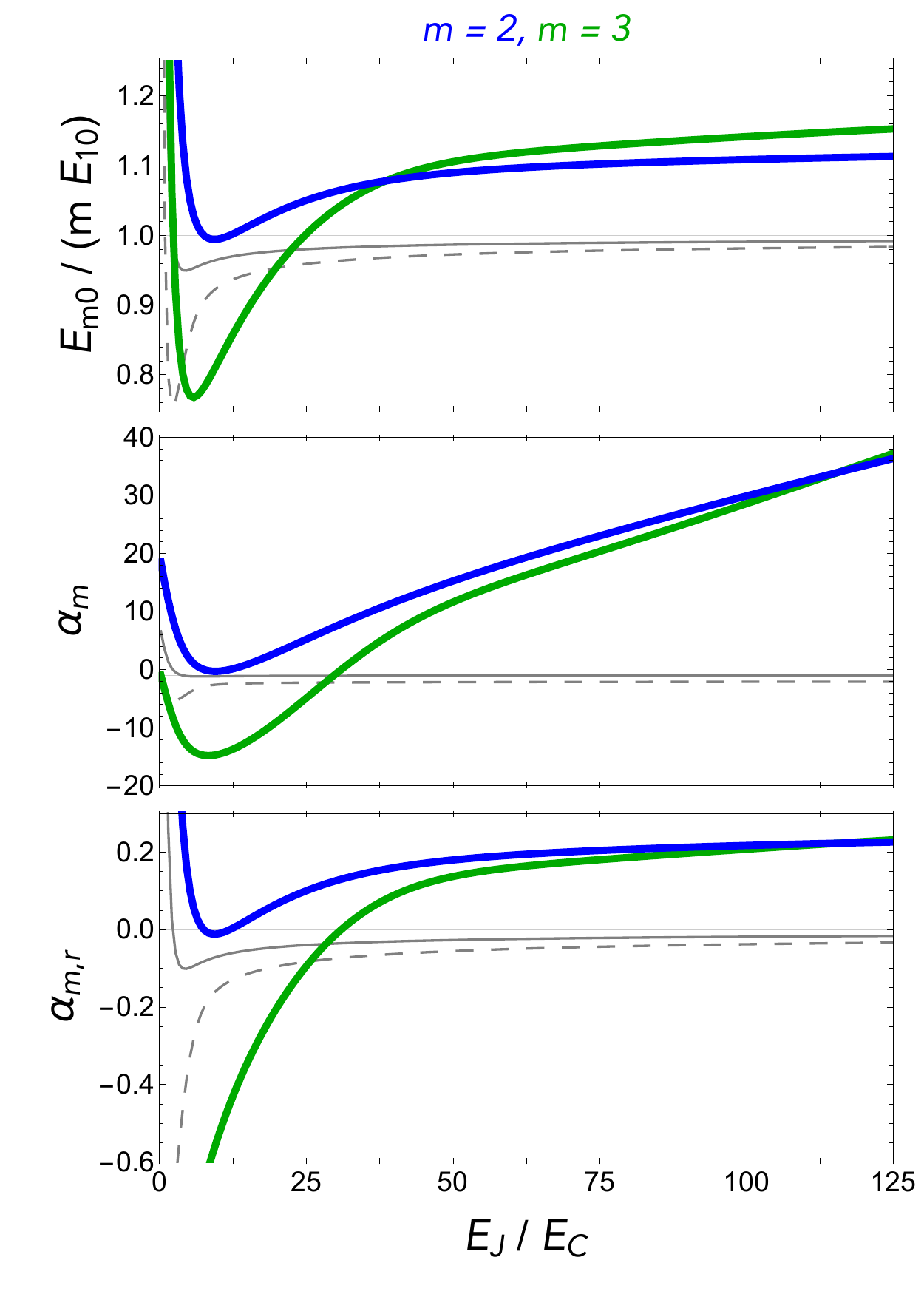}
\hspace{2em}
\includegraphics[width=.29\textwidth]{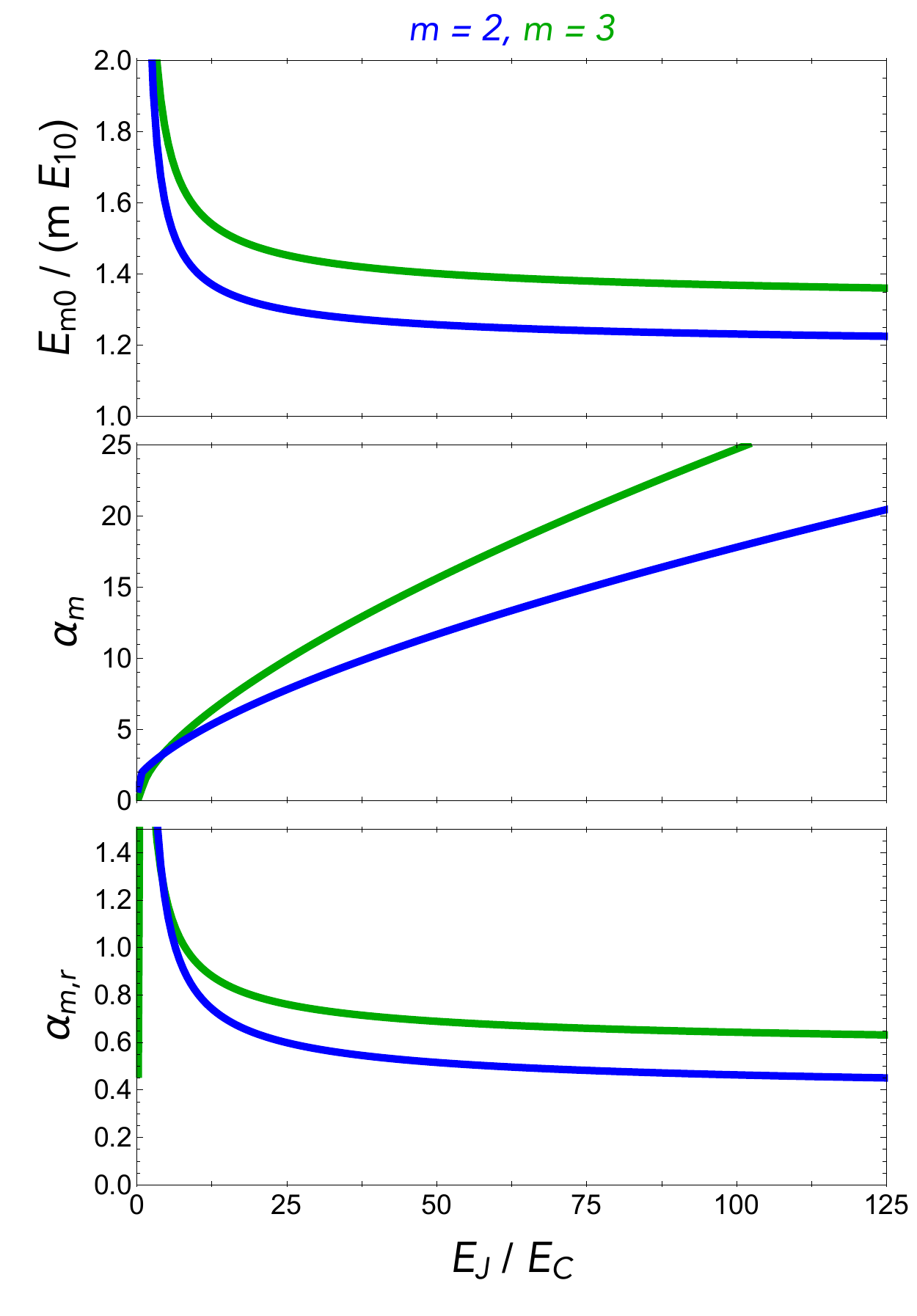}
\caption{
Relative energy ratio $E_{m0}/(mE_{10})$ for the second and third levels of the FEmon, together with the absolute and relative anharmonicity parameters $\alpha$ and $\alpha_r$ as a function of $E_J/E_C$. (a) and (b) correspond to $a = 0$ and $a/b = -5$ respectively. The gray curves in (a) correspond to the standard transmon ($m=2$ solid, $m=3$ dashed). In both cases, (a) and (b), the curves show a well defined asymptotic behavior for large $E_J/E_C$. However, this behavior is approached from below in the incipient ferrroelectric regime and from above in the ferroelectric one. 
Furthermore, the $\alpha$ and $\alpha_r$ curves demonstrate that the FEmon combines charge-noise insensitiveness with strong anharmonicity.
}
\label{f:anharmonicity}
\end{figure*}

To further assess this optimization, we compute the energy-level differences $E_{m0} = E_m - E_0$ at offset charge $n_g = 1/2$ as a function of the ratio $E_J/E_C$. The results, illustrated in Fig.~\ref{f:anharmonicity} (a) for $a = 0$, show that the spacing between adjacent FEmon levels in the incipient ferroelectric regime ultimately scales as $E_C (E_J/E_C)^{2/3}$. 
This should be contrasted with the $E_C (E_J/E_C)^{1/2}$ scaling far away from ferroelectric criticality, which is characteristic of the standard transmon~\cite{schoelkopf07-pra}.  

To further quantify the distinct anharmonicity of the FEmon, we follow Ref.~\cite{schoelkopf07-pra} and define the absolute and relative anharmonicity from the transition energies $E_{mm-1}$ as
$\alpha_m = \frac{E_{mm-1} - E_{10}}{4 E_C}$, 
$\alpha_{m,r} = \frac{E_{mm-1} - E_{10}}{E_{10}}$.
These measures are compared with the standard $b=0$ transmon model in Fig.~\ref{f:anharmonicity} (a). As can be seen, once the charge dispersion is sufficiently suppressed ($E_J/E_C \gg 1$), the FEmon anharmonicity continues to increase with increasing $E_J/E_C$. In fact, unlike the standard transmon, the FEmon anharmonicity retains a weak dependence on $E_J/E_C$, increasing monotonically with no sign of saturation.

\section{Ferroelectric regime}

We now consider the FEmon operating in its ferroelectric regime, where the distinct behavior in the charge sector further enhances the features already obtained in the incipient ferroelectric regime. 
In the ferroelectric regime $a<0$ and the system develops a spontaneous charge offset
$\langle \hat n \rangle = n_0 = \pm\sqrt{|a|/b}$ associated with the ferroelectric polarization. The two possible signs of the polarization correspond to two equivalent operating configurations.

\begin{figure*}[t!]
\begin{tabular}{r c l}
\vspace{-43em}\\
\footnotesize $\displaystyle {a\over b} $ & $=$ & $-0.5$ \\
\vspace{7em}\\
 \footnotesize $\displaystyle {a\over b} $    & $=$ & $-1$\\ 
\vspace{7em}\\
\footnotesize $\displaystyle {a\over b} $ & $=$ & $-5$\\
\end{tabular}
\includegraphics[width=.9\textwidth]{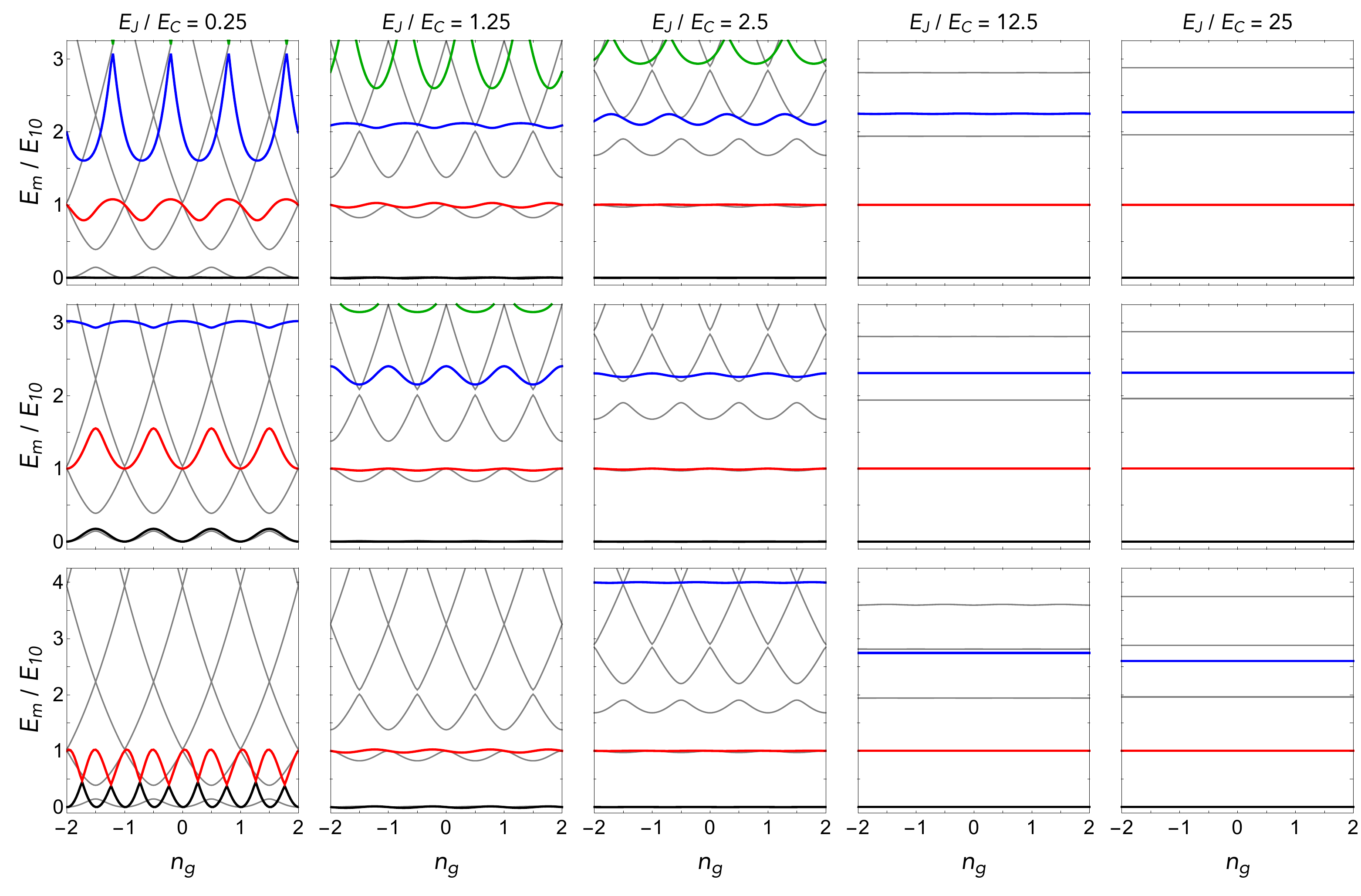} 
\caption{
Low-lying energy levels $E_m$ ($m=0,1,2,3$, shown in black, red, blue, and green, respectively) of the FEmon in the ferroelectric regime, compared with the corresponding transmon levels (gray), as a function of the effective offset charge $n_g$ for different values of $E_J/E_C$ and $a/b$. 
The spontaneous offset is chosen as $n_0>0$, and the energies are expressed in units of the transition energy $E_{10}$ evaluated at $n_g=0$.
\label{f:levels.fe}}
\end{figure*}

Figure~\ref{f:levels.fe} shows the low-lying spectrum of the FEmon in this regime. The corresponding charging energy is defined as
$E_C=\frac{a}{2}+\frac{b}{4}+b\left(1+\frac{3}{2}|n_0|\right)|n_0|$.
As in the standard transmon and in the incipient ferroelectric regime, increasing the ratio $E_J/E_C$ leads to band flattening and therefore to a strong suppression of charge-noise sensitivity. In the ferroelectric regime, however, this flattening is remarkably more effective and already occurs at moderate values of $E_J/E_C$, as illustrated by the $E_J/E_C=1.25$ curves.

At the same time, the suppression of charge sensitivity is accompanied by a substantial enhancement of the qubit anharmonicity. Owing to the increasingly dispersionless character of the spectrum, this enhancement is not restricted to special operating points such as $n_g=1/2$ or $n_g=0$, but extends over the full range of $n_g$. Moreover, the anharmonicity increases rapidly as the capacitor is driven deeper into the ferroelectric regime, i.e., with increasing $|a|$ for $a<0$.

These optimized features are quantified in Fig.~\ref{f:anharmonicity} (b). In particular, the evolution of the relative energy ratio $E_{m0}/(mE_{10})$ with increasing $E_J/E_C$ shows that, as in the incipient ferroelectric regime, the FEmon develops a well-defined asymptotic behavior in the charge-noise-insensitive regime. In contrast to the incipient case, however, this the asymptotic behavior is now approached from above, reflecting the qualitatively distinct spectral renormalization induced by the ferroelectric charge response.

This distinction is even more pronounced in the behavior of the absolute and relative anharmonicity parameters, $\alpha$ and $\alpha_r$. As shown in Fig.~\ref{f:anharmonicity} (b), both quantities remain substantial over a broad range of $E_J/E_C$, including in the regime where the spectrum has already become dispersionless. This demonstrates that, unlike in the conventional transmon where charge-noise suppression is typically accompanied by a marked reduction of anharmonicity, the FEmon preserves strong level selectivity while remaining robust against charge fluctuations.

Altogether, these distinct features provide additional flexibility for device optimization and tuning. In particular, the ferroelectric regime of the FEmon offers the optimal tradeoff between charge-noise insensitivity and enhanced anharmonicity. 
Importantly, we note that these features enable such optimization without requiring particularly large values of $E_J$ itself.

\section{Discussion and Perspectives}

We have introduced the concept of the ferroelectric transmon, denoted as FEmon. This new superconducting qubit exploits the intrinsically nonlinear response of ferroelectric materials to simultaneously optimize two key performance metrics: qubit anharmonicity and insensitivity to charge noise.

As in the standard transmon, the suppression of charge-noise sensitivity can be understood from the effective irrelevance of the periodic boundary condition on the qubit wave function,
$\psi(\phi)=\psi(\phi+2\pi)$,
i.e., of the underlying $U(1)$ symmetry encoded in the Josephson potential $-E_J\cos\phi$~\cite{schoelkopf07-pra}. For states strongly localized around $\phi=0$, the influence of adjacent Josephson wells is reduced to rare $2\pi$-phase-slip events, which are exponentially suppressed. In this limit, a translation in $\hat n$ can be effectively gauged out, and charge-noise insensitivity is therefore likewise ensured whenever $E_J/E_C\gg1$. In the FEmon, however, the possibility of reaching this regime depends on the independent parameters $a$, $b$, and $E_J$, thereby opening additional routes to combine charge-noise protection with enhanced anharmonicity.

A particularly distinctive feature emerges in the ferroelectric regime, where the charge sector acquires a double-well structure. This gives rise to a new effective energy scale
$E_{C}^{\mathrm{FE}}\sim\sqrt{|a|E_J}$. 
As the well depth increases, the lowest eigenstates become progressively confined within the two local minima. The formation of the first pair of confined states is governed by the interplay between the charging and Josephson energies and, importantly, does not require particularly large values of $E_J$. 
Well inside the ferroelectric regime, the two wells can be approximated as quasi-harmonic. Then the splitting of the energy levels is ultimately determined by quantum tunneling through the barrier and can be estimated within the WKB approximation.
In this regime, the FEmon anharmonicity becomes exponentially enhanced, with
$\frac{E_{21}}{E_{10}}
\sim
{\sqrt{\pi} \over 4}
\exp\Big(
\frac{a^{2}}{2^{1/3}b\sqrt{|a|E_J}}
\Big)$.
This exponential scaling establishes a fundamentally new operating regime for superconducting qubitss.

Materials relevant for the realization of the FEmon include perovskites such as BaTiO$_3$, simple binary oxides such as HfO$_2$, and AlScN~\cite{setter2006ferroelectric,park2018review,schopeder21}. Importantly, the enhanced performance of the FEmon relies primarily on the nonlinear ferroelectric response rather than on the existence of spontaneous, switchable polarization itself. Consequently, quantum paraelectrics and incipient ferroelectrics, such as SrTiO$_3$~\cite{chandra2017prospects}, also constitute promising material platforms.

Moreover, by exploiting the ferroelectric softness---i.e., the smallness of the associated charging energy---the desired operating regime can be achieved with substantially reduced critical currents in the Josephson junction. This feature additionally broadens the range of superconducting materials and device architectures that may be employed in superconducting quantum technologies.

These perspectives are expected to motivate not only further fundamental studies of nonlinear superconducting qubits but also new avenues for quantum engineering and technological development.

\bibliography{bib.bib}

\end{document}